\documentstyle[12pt]{article}
\input psfig

\newcommand{\be}{\begin{equation}}
\newcommand{\ee}{\end{equation}}
\newcommand{\bea}{\begin{eqnarray}}
\newcommand{\eea}{\end{eqnarray}}

\newcommand{\half}{{\scriptstyle{{1\over 2}}}}

\newcommand{\tfour}{{\scriptstyle{{3\over 4}}}}

\newcommand{\third}{{\scriptstyle{{1\over 3}}}}

\newcommand{\quart}{{\scriptstyle{{1\over 4}}}}
\newcommand{\real}{\relax{\rm I\kern-.18em R}}
\newcommand{\zahlen}{{\rm Z \!\! Z}}

\newcommand{\Tr}{\mbox{\,Tr\,}}

\newcommand{\cO}{{\cal O}}
\newcommand{\cE}{{\cal E}}

\newcommand{\veps}{\varepsilon}
\newcommand{\fie}{\varphi}

\newcommand{\integer}{\relax{\rm I\kern-.18em N}}

\newcommand{\cA}{{\cal A}}

\newcommand{\basispl}{
   \put(-.5,-.5){\line(1,0){1}}
   \put(.5,-.5){\line(0,1){1}}
   \put(.5,.5){\line(-1,0){1}}
   \put(-.5,.5){\line(0,-1){1}}}
\newcommand{\basisar}{
   \put(0,-.5){\vector(1,0){0}}
   \put(.5,0){\vector(0,1){0}}
   \put(0,.5){\vector(-1,0){0}}
   \put(-.5,0){\vector(0,-1){0}}}
\newcommand{\plaq}{\setlength{\unitlength}{.5cm}\raisebox{-.2cm}{
   \begin{picture}(1.2,1.2)(-.6,-.6)
   \basispl\basisar
   \put(-.5,-.5){\circle*{.2}}
   \put(-.60,-.60){\makebox(0,0)[tr]{\footnotesize $x$}}
   \put(-.55,0.4){\makebox(0,0)[r]{\footnotesize $\nu$}}
   \put(0.4,-.60){\makebox(0,0)[t]{\footnotesize $\mu$}}
   \end{picture}}}
\newcommand{\wplaqone}{\setlength{\unitlength}{1cm}\raisebox{-.5cm}{
   \begin{picture}(1.2,1.2)(-.6,-.6)
   \put(.25,-.5){\line(0,1){1}}
   \put(.25,.5){\line(-1,0){.5}}
   \put(-.25,.5){\line(0,-1){1}}
   \put(-.25,0){\line(1,0){.1}}
   \put(-.10,0){\line(1,0){.1}}
   \put(.05,0){\line(1,0){.1}}
   \put(.2,0){\line(1,0){.05}}
   \put(-.25,-.5){\circle*{.1}}
   \put(-.25,.5){\circle*{.1}}
   \put(.25,-.5){\circle*{.1}}
   \put(.25,.5){\circle*{.1}}
   \put(-.25,0){\circle*{.1}}
   \put(.25,0){\circle*{.1}}
   \put(-.25,-.15){\vector(0,1){0}}
   \put(-.25,.35){\vector(0,1){0}}
   \put(.1,.5){\vector(1,0){0}}
   \put(.25,.15){\vector(0,-1){0}}
   \put(.25,-.35){\vector(0,-1){0}}
   \put(-.0,-.55){\makebox(0,0)[tr]{\footnotesize $x$}}
   \put(-.34,-.3){\makebox(0,0)[r]{\footnotesize $a$}}
   \put(-.43,.3){\makebox(0,0)[t]{\footnotesize $b$}}
   \put(0,.8){\makebox(0,0)[t]{\footnotesize $\mu$}}
   \end{picture}}}
\newcommand{\wplaqonem}{\setlength{\unitlength}{1cm}\raisebox{-.5cm}{
   \begin{picture}(1.2,1.2)(-.6,-.6)
   \put(-.29,-.5){\line(0,1){.54}}
   \put(-.29,.04){\line(1,0){.54}}
   \put(.25,.04){\line(0,1){.46}}
   \put(.25,.5){\line(-1,0){.46}}
   \put(-.21,.5){\line(0,-1){.40}}
   \put(-.21,-.04){\line(1,0){.46}}
   \put(.25,-.04){\line(0,-1){.46}}
   \put(-.25,-.5){\circle*{.1}}
   \put(-.21,.5){\circle*{.1}}
   \put(.25,.5){\circle*{.1}}
   \put(-.29,-.5){\circle*{.1}}
   \put(.25,-.5){\circle*{.1}}
   \put(-.29,-.15){\vector(0,1){0}}
   \put(-.21,.25){\vector(0,-1){0}}
   \put(-.1,.5){\vector(-1,0){0}}
   \put(.2,-.04){\vector(1,0){0}}
   \put(.1,.04){\vector(1,0){0}}
   \put(.25,.35){\vector(0,1){0}}
   \put(.25,-.35){\vector(0,-1){0}}
   \put(-.30,-.60){\makebox(0,0)[tr]{\footnotesize $x$}}
   \put(-.38,-.3){\makebox(0,0)[r]{\footnotesize $a$}}
   \put(.4,.45){\makebox(0,0)[t]{\footnotesize $b$}}
   \put(0,.3){\makebox(0,0)[t]{\footnotesize $\mu$}}
   \end{picture}}}
\newcommand{\wplaqtwo}{\setlength{\unitlength}{1cm}\raisebox{-.5cm}{
   \begin{picture}(1.2,1.2)(-.6,-.6)
   \put(-.5,-.25){\line(0,1){.5}}
   \put(-.5,.25){\line(1,0){1}}
   \put(.5,.25){\line(0,-1){.5}}
   \put(.5,-.25){\line(-1,0){.5}}
   \put(0,-.25){\line(0,1){.1}}
   \put(0,-.10){\line(0,1){.1}}
   \put(0,.05){\line(0,1){.1}}
   \put(0,.20){\line(0,1){.1}}
   \put(-.5,-.25){\circle*{.1}}
   \put(0,-.25){\circle*{.1}}
   \put(.5,-.25){\circle*{.1}}
   \put(-.5,.25){\circle*{.1}}
   \put(0,.25){\circle*{.1}}
   \put(-.5,.25){\circle*{.1}}
   \put(-.5,.1){\vector(0,1){0}}
   \put(-.15,.25){\vector(1,0){0}}
   \put(.35,.25){\vector(1,0){0}}
   \put(.5,-.1){\vector(0,-1){0}}
   \put(.2,-.25){\vector(-1,0){0}}
   \put(0,-.3){\makebox(0,0)[tr]{\footnotesize $x$}}
   \put(.35,.42){\makebox(0,0)[r]{\footnotesize $a$}}
   \put(-.65,.15){\makebox(0,0)[t]{\footnotesize $b$}}
   \put(-.25,.5){\makebox(0,0)[t]{\footnotesize $\mu$}}
   \end{picture}}}
\newcommand{\wplaqtwom}{\setlength{\unitlength}{1cm}\raisebox{-.5cm}{
   \begin{picture}(1.2,1.2)(-.6,-.6)
   \put(-.5,-.25){\line(0,1){.54}}
   \put(-.5,.29){\line(1,0){.54}}
   \put(.5,.21){\line(-1,0){.40}}
   \put(.5,.21){\line(0,-1){.46}}
   \put(.5,-.25){\line(-1,0){.46}}
   \put(.04,.29){\line(0,-1){.54}}
   \put(-.04,.21){\line(0,-1){.46}}
   \put(-.5,-.25){\circle*{.1}}
   \put(.5,-.25){\circle*{.1}}
   \put(-.5,.29){\circle*{.1}}
   \put(.5,.21){\circle*{.1}}
   \put(-.5,.1){\vector(0,1){0}}
   \put(-.15,.29){\vector(1,0){0}}
   \put(.2,.21){\vector(-1,0){0}}
   \put(.04,-.05){\vector(0,-1){0}}
   \put(-.04,-.15){\vector(0,-1){0}}
   \put(.5,.1){\vector(0,1){0}}
   \put(.35,-.25){\vector(1,0){0}}
   \put(-.55,-.3){\makebox(0,0)[tr]{\footnotesize $x$}}
   \put(.35,-.42){\makebox(0,0)[r]{\footnotesize $a$}}
   \put(-.65,.15){\makebox(0,0)[t]{\footnotesize $b$}}
   \put(-.25,.54){\makebox(0,0)[t]{\footnotesize $\mu$}}
   \end{picture}}}
\newcommand{\wplaqthree}{\setlength{\unitlength}{1cm}\raisebox{-.5cm}{
   \begin{picture}(1.2,1.2)(-.6,-.6)
   \put(-.5,-.25){\line(0,1){.5}}
   \put(-.5,.25){\line(1,0){1}}
   \put(.5,.25){\line(0,-1){.5}}
   \put(0,-.25){\line(-1,0){.5}}
   \put(0,-.25){\line(0,1){.1}}
   \put(0,-.10){\line(0,1){.1}}
   \put(0,.05){\line(0,1){.1}}
   \put(0,.20){\line(0,1){.1}}
   \put(-.5,-.25){\circle*{.1}}
   \put(0,-.25){\circle*{.1}}
   \put(.5,-.25){\circle*{.1}}
   \put(-.5,.25){\circle*{.1}}
   \put(0,.25){\circle*{.1}}
   \put(-.5,.25){\circle*{.1}}
   \put(-.5,.1){\vector(0,1){0}}
   \put(.35,.25){\vector(1,0){0}}
   \put(-.15,.25){\vector(1,0){0}}
   \put(.5,-.05){\vector(0,-1){0}}
   \put(-.35,-.25){\vector(-1,0){0}}
   \put(.1,-.35){\makebox(0,0)[tr]{\footnotesize $x$}}
   \put(.37,.42){\makebox(0,0)[r]{\footnotesize $\mu$}}
   \put(-.65,.15){\makebox(0,0)[t]{\footnotesize $b$}}
   \put(-.25,.5){\makebox(0,0)[t]{\footnotesize $a$}}
   \end{picture}}}
\newcommand{\wplaqthreem}{\setlength{\unitlength}{1cm}\raisebox{-.5cm}{
   \begin{picture}(1.2,1.2)(-.6,-.6)
   \put(.5,-.25){\line(0,1){.54}}
   \put(.5,.29){\line(-1,0){.54}}
   \put(-.5,.21){\line(1,0){.40}}
   \put(-.5,.21){\line(0,-1){.46}}
   \put(-.5,-.25){\line(1,0){.46}}
   \put(-.04,.29){\line(0,-1){.54}}
   \put(.04,.21){\line(0,-1){.46}}
   \put(.5,-.25){\circle*{.1}}
   \put(-.5,-.25){\circle*{.1}}
   \put(.5,.29){\circle*{.1}}
   \put(-.5,.21){\circle*{.1}}
   \put(.5,-.1){\vector(0,-1){0}}
   \put(.35,.29){\vector(1,0){0}}
   \put(-.35,.21){\vector(-1,0){0}}
   \put(-.04,.05){\vector(0,1){0}}
   \put(.04,.2){\vector(0,1){0}}
   \put(-.5,-.1){\vector(0,-1){0}}
   \put(-.15,-.25){\vector(1,0){0}}
   \put(.15,-.35){\makebox(0,0)[tr]{\footnotesize $x$}}
   \put(-.2,-.42){\makebox(0,0)[r]{\footnotesize $a$}}
   \put(-.15,.15){\makebox(0,0)[t]{\footnotesize $b$}}
   \put(.25,.54){\makebox(0,0)[t]{\footnotesize $\mu$}}
   \end{picture}}}
\newcommand{\stapup}{\setlength{\unitlength}{.5cm}\raisebox{-.2cm}{
   \begin{picture}(1.2,1.2)(-.6,-.6)
   \put(.5,-.5){\line(0,1){1}}
   \put(.5,.5){\line(-1,0){1}}
   \put(-.5,.5){\line(0,-1){1}}
   \put(.5,0){\vector(0,-1){0}}
   \put(0,.5){\vector(1,0){0}}
   \put(-.5,0){\vector(0,1){0}}
   \put(-.5,-.5){\circle*{.2}}
   \put(-.45,-.65){\makebox(0,0)[tr]{\footnotesize $x$}}
   \put(-.55,0){\makebox(0,0)[r]{\footnotesize $\nu$}}
   \put(0.1,.70){\makebox(0,0)[b]{\footnotesize $\mu$}}
   \end{picture}}}
\newcommand{\linka}{\setlength{\unitlength}{.5cm}\raisebox{-.2cm}{
   \begin{picture}(0.4,0.6)(-.6,-.6)
   \put(-.5,.5){\line(0,-1){1}}
   \put(-.5,0){\vector(0,1){0}}
   \put(-.5,-.5){\circle*{.2}}
   \put(-.60,-.60){\makebox(0,0)[tr]{\footnotesize $x$}}
   \put(-.55,0.3){\makebox(0,0)[r]{\footnotesize $a$}}
   \end{picture}}}
\newcommand{\linkap}{\setlength{\unitlength}{.5cm}\raisebox{-.2cm}{
   \begin{picture}(0.6,0.6)(-.6,-.6)
   \put(-.5,.5){\line(0,-1){1}}
   \put(-.5,0){\vector(0,1){0}}
   \put(.20,-.60){\makebox(0,0)[tr]{\footnotesize $x\!+\!\hat\mu$}}
   \end{picture}}}
\newcommand{\linkb}{\setlength{\unitlength}{.5cm}\raisebox{-.2cm}{
   \begin{picture}(0.4,0.6)(-.6,-.6)
   \put(-.5,.5){\line(0,-1){1}}
   \put(-.5,0){\vector(0,-1){.2}}
   \put(-.5,-.5){\circle*{.2}}
   \put(-.60,-.60){\makebox(0,0)[tr]{\footnotesize $x$}}
   \put(-.55,0.3){\makebox(0,0)[r]{\footnotesize $a$}}
   \end{picture}}}
\newcommand{\linkbp}{\setlength{\unitlength}{.5cm}\raisebox{-.2cm}{
   \begin{picture}(0.6,0.6)(-.6,-.6)
   \put(-.5,.5){\line(0,-1){1}}
   \put(-.5,0){\vector(0,-1){.2}}
   \put(.20,-.60){\makebox(0,0)[tr]{\footnotesize $x\!+\!\hat\mu$}}
   \end{picture}}}
\newcommand{\linkhmu}{\setlength{\unitlength}{.5cm}\raisebox{-.2cm}{
   \begin{picture}(1.2,1.2)(-.6,-.6)
   \put(.5,0){\line(-1,0){1}}
   \put(0,0){\vector(1,0){0.1}}
   \put(-.5,0){\circle*{.2}}
   \put(-.35,-.25){\makebox(0,0)[tr]{\footnotesize $x$}}
   \put(0.4,-.3){\makebox(0,0)[r]{\footnotesize $\mu$}}
   \end{picture}}}

\newcommand{\stapupa}{\setlength{\unitlength}{.5cm}\raisebox{-.2cm}{
   \begin{picture}(1.2,1.2)(-.6,-.6)
   \put(.5,-.5){\line(0,1){1}}
   \put(.5,.5){\line(-1,0){1}}
   \put(-.5,.5){\line(0,-1){1}}
   \put(.5,0){\vector(0,-1){0}}
   \put(0,.5){\vector(1,0){0}}
   \put(-.5,0){\vector(0,1){0}}
   \put(-.5,-.5){\circle*{.2}}
   \put(-.60,-.60){\makebox(0,0)[tr]{\footnotesize $x$}}
   \put(0,0){\makebox(0,0)[r]{\footnotesize $a$}}
   \put(0.1,.7){\makebox(0,0)[b]{\footnotesize $\mu$}}
   \end{picture}}}
\newcommand{\stapupap}{\setlength{\unitlength}{.5cm}\raisebox{-.2cm}{
   \begin{picture}(1.2,1.2)(-.6,-.6)
   \put(.5,-.5){\line(0,1){1}}
   \put(-.5,.5){\line(1,0){.2}}
   \put(-.20,.5){\line(1,0){.2}}
   \put(.1,.5){\line(1,0){.2}}
   \put(.4,.5){\line(1,0){.1}}
   \put(-.5,.5){\line(0,-1){1}}
   \put(.5,0){\vector(0,-1){0}}
   \put(-.5,0){\vector(0,1){0}}
   \put(-.5,-.5){\circle*{.2}}
   \put(-.60,-.60){\makebox(0,0)[tr]{\footnotesize $x$}}
   \put(0,0){\makebox(0,0)[r]{\footnotesize $a$}}
   \put(0.1,.8){\makebox(0,0)[b]{\footnotesize $\mu$}}
   \end{picture}}}
\newcommand{\stapupaq}{\setlength{\unitlength}{.5cm}\raisebox{-.2cm}{
   \begin{picture}(1.2,1.2)(-.6,-.6)
   \put(.5,-.5){\line(0,1){1.1}}
   \put(.5,.6){\line(-1,0){.85}}
   \put(-.5,.45){\line(1,0){.85}}
   \put(-.5,.45){\line(0,-1){.95}}
   \put(.5,0){\vector(0,-1){0}}
   \put(0.4,.62){\vector(1,0){0}}
   \put(0.1,.45){\vector(1,0){0}}
   \put(-.5,0){\vector(0,1){0}}
   \put(-.5,-.5){\circle*{.2}}
   \put(-.60,-.60){\makebox(0,0)[tr]{\footnotesize $x$}}
   \put(0,0){\makebox(0,0)[r]{\footnotesize $a$}}
   \put(0.1,.85){\makebox(0,0)[b]{\footnotesize $\mu$}}
   \end{picture}}}
\newcommand{\stapdw}{\setlength{\unitlength}{.5cm}\raisebox{-.2cm}{
   \begin{picture}(1.2,1.2)(-.6,-.6)
   \put(.5,-.5){\line(0,1){1}}
   \put(.5,-.5){\line(-1,0){1}}
   \put(-.5,.5){\line(0,-1){1}}
   \put(.5,0){\vector(0,1){0}}
   \put(0,-.5){\vector(1,0){0}}
   \put(-.5,0){\vector(0,-1){0}}
   \put(-.5,.5){\circle*{.2}}
   \put(-.60,.85){\makebox(0,0)[tr]{\footnotesize $x$}}
   \put(-.55,0){\makebox(0,0)[r]{\footnotesize $\nu$}}
   \put(0,-.60){\makebox(0,0)[t]{\footnotesize $\mu$}}
   \end{picture}}}

\newcommand{\AmS}{{\protect\the\textfont2
  A\kern-.1667em\lower.5ex\hbox{M}\kern-.125emS}}

\begin{document}
\vskip-1cm
\hfill INLO-PUB-20/95
\vskip5mm
\begin{center}
{\LARGE{\bf{\underline{The electroweak sphaleron on the lattice}}}
}\\
\vspace{1cm}
{\large Margarita Garc\'{\i}a P\'erez and Pierre van Baal
} \\
\vspace{1cm}
Instituut-Lorentz for Theoretical Physics,\\
University of Leiden, PO Box 9506,\\
NL-2300 RA Leiden, The Netherlands.\\
\end{center}
\vspace*{5mm}{\narrower\narrower{\noindent
\underline{Abstract:}
We study the properties of the electroweak sphaleron on a finite lattice.
The cooling algorithm for saddle points is used to obtain the static classical
solutions of the SU(2)-Higgs field theory. Results are presented for $M_H=
\infty,\,M_W,\,\tfour M_W$. After performing finite size scaling we find good
agreement with the results obtained from variational approaches. Of relevance
for numerical determinations of the transition rate is that the lattice
artefacts are surprisingly small for $M_W\approx M_H$.}\par}

\section{Introduction}

In this paper we will study the sphaleron solutions for the SU(2)-Higgs field
theory, using the lattice approximation and an algorithm to find saddle-point
solutions. The sphaleron is a solution of the static equations of motion, i.e.
a stationary point of the energy functional, which has precisely one unstable
direction. This direction corresponds to the tunnelling path associated to
the (approximate) instanton. Due to the spherical symmetry, variational
analysis using a radial ansatz has provided accurate results quite some time
ago~\cite{dashen,klin}. However, due to the recent interest of studying the
sphaleron transition rates on a lattice~\cite{amb}, the question arises how
big the lattice artefacts are for the particular sizes of lattices that are
employed in the numerical analysis. The lattice destroys the rotational
invariance and a variational analysis does no longer seem very practical.
Furthermore, in the absence of rotational symmetry in the continuum, the
method discussed can be used with the same ease.

We have reported earlier~\cite{lat94} on the sphaleron solutions where the
length of the Higgs field is frozen. In the unitary gauge this means that we
only need to consider gauge degrees of freedom.
We recall that above $M_H=12M_W$ the sphaleron undergoes a series of
bifurcations~\cite{yaf}, acquiring at each bifurcation an additional negative
mode, while new solutions, so-called deformed sphalerons split off. For
infinite $M_H$, where the model is identical to the gauged non-linear
sigma model, there is an infinite number of solutions ranging in energy from
$5.41M_W/\alpha_W$ to the energy of the lowest deformed sphaleron
$5.07M_W/\alpha_W$, which has only one negative mode (the number of unstable
modes increases with increasing energy). These solutions are related to the
electroweak skyrmions~\cite{eilam}.

Here we will include the scalar field in the analysis to allow study of the
electroweak sphalerons (at $\theta_W=0$) for a more interesting range of
parameters. We will report results for $M_H=M_W$ and $M_H=\tfour M_W$, the
latter value corresponding to $M_H\approx 60$GeV, the present experimental
bound for the Higgs mass~\cite{pdb}. Since for
finite values of the Higgs self-coupling the scalar field is allowed to vanish
at the center, these solutions are smoother (have smaller lattice artefacts)
than for the electroweak skyrmions. We first present the new algorithm to find
the extremum of the energy functional, based on minimizing the square of the
equations of motion.  A careful analysis of the finite size scaling is
performed, to allow for a reliable extrapolation to the infinite volume limit.
The results agree accurately with those obtained from the variational analysis.
For $M_H=M_W$ the lattice artefacts are to a good degree described by the
formula $E=E_0-0.3 (aM_W)^2-0.3 (aM_W)^4$, whereas the volume corrections
are described by $3.641+18.1(M_WL)^{-1}e^{-M_WL}$ (the infinite volume
variational result~\cite{yaf} is 3.6417) all in units of $M_W/\alpha_W$,
where $\alpha_W=g^2/4\pi$ is the electroweak fine-structure constant.

\section{The model}

The dynamical variables for the SU(2)-Higgs model on the lattice are the
gauge group variables $V_\mu(x)$, defined on the link that runs from $x$ to
$x+\hat\mu$, and the Higgs field in the fundamental representation of
SU(2) (a complex two-component spinor) defined on the site $x$. This Higgs
field can be represented by its length $\rho(x)$ (in the continuum this
neutral Higgs field will be denoted by $\phi(x)$)
and a SU(2) matrix $\sigma(x)$, which is associated to the gauge degree of
freedom and can be reabsorbed into the links via the change of
variables~\cite{lan}
\be
U_\mu(x) = \sigma(x) V_\mu(x) \sigma(x+\mu).
\ee
This gives the Higgs model in the unitary gauge.
The lattice action is ($U_\mu(x)=\linkhmu$)
\be
S\!=\!\frac{a^n}{g^2a^4}\!\sum_x\!\left\{\sum_{\mu,\nu}\!\!\Tr\!\left(1\!\!-
\!\plaq\right)\!-\!\kappa\!\!\sum_{\mu}\!\rho(x)\rho(x\!+\!\hat\mu)\Tr\!\left(
U_\mu(x)\right)\!+\!\rho^2(x)\!+\!\lambda(\rho^2(x)\!-\!1)^2\!-\!C_0
\right\}
.\label{eq:Swil}\\
\ee
For $n=4$ ($n=3$) the continuum action (energy) functional is recovered by
rescaling the fields and coupling constants. Introducing a lattice spacing
$a$, to convert to dimensionful parameters, one first scales the fields
to get the correct normalizations for the kinetic terms.
\be
U_\mu(x)=\exp(aA_\mu(x)),\quad A_\mu(x)=-igA_\mu^a(x)\frac{\tau_a}{2},
\quad\rho^2(x)=\frac{a^2g^2}{2\kappa}\phi^2(x),
\ee
where $\tau_a$ are the Pauli matrices. The
continuum parameters $M_W$, $M_H$ and $\bar\lambda$
are given by
\be
\bar\lambda=\frac{g^2\lambda}{4\kappa^2},\quad
(aM_W)^2=\frac{\kappa v^2}{2},\quad(aM_H)^2=\frac{4\lambda v^2}{\kappa},
\ee
with $v$ the lattice vacuum expectation value
\be
v^2=\frac{8\kappa+2\lambda-1}{2\lambda}.
\ee
Introducing the parameters
\be
\bar\kappa\equiv 2M_W^2a^2,\quad r\equiv M_H/M_W,
\ee
one can eliminate $\lambda$ and $\kappa$ in favour of these more physical
parameters
\be
\lambda=\frac{r^2\kappa^2}{8},\quad\kappa=\frac{-(32-\bar\kappa r^2)+
\sqrt{(32-\bar\kappa r^2)^2+16 r^2}}{2r^2}.
\ee
Note that for $r\rightarrow\infty$, $v\rightarrow 1$ and
$\kappa\rightarrow\bar\kappa$.

In this paper we are interested in the energy functional, with $U_0(x)=I$,
and all fields time independent. Note that restricting the sums over the
indices to three dimensions leaves an extra term $-2\kappa\rho^2(x)$ from
the time component of the hopping term. We have chosen our conventions such
that the gauge coupling constant can be factored out, allowing us to express
the energies in units of $M_W/\alpha_W$
\bea
\cE&=&\frac{M_W}{2\pi\alpha_W\sqrt{2\bar\kappa}}\!\sum_x\!\Bigl\{
\sum_{\mu,\nu}\!\!\Tr\!\left(1\!\!-\!\plaq\right)\!-\!\kappa\!\!\sum_{\mu}
\!\rho(x)\rho(x\! +\!\hat\mu)\Tr\!\left(U_\mu(x)\right)\!+\!(1\!-\!2\kappa)
\rho^2(x)\nonumber\\&&\hspace{7cm}+\lambda(\rho^2(x)\!-\!1)^2\!\!-\!C_0
\Bigr\}.\label{eq:Esphal}
\eea
{}From now on all indices are assumed to run over the values 1-3.
The constant $C_0$ normalizes the vacuum ($U_\mu(x)\equiv I$ and
$\rho(x)\equiv v$) energy to zero,
\be
C_0=(1\!-\!8\kappa)v^2\!+\!\lambda(v^2\!-\!1)^2=\lambda\!-\!\half(aM_H)^2
(aM_W)^2.
\ee
For ease of reference we quote the continuum expression for the energy
functional in the unitary gauge using our conventions ($F_{\mu\nu}\!=\!
\partial_\mu A_\nu\!-\!\partial_\nu A_\mu\!+\![A_\mu,A_\nu]\equiv
-igF_{\mu\nu}^a\tau_a/2$)
\be
\cE=\frac{1}{2g^2}\!\int d^3x|\Tr(F_{\mu\nu})^2|\!+\!\int d^3x \left[\half(
\partial_\mu\phi)^2\!-\!\quart\Tr(A^2_\mu)\phi^2\!+\!\bar\lambda\left(\phi^2\!
-\!M_H^2/8\bar\lambda\right)^2\right],
\ee

\section{Cooling}
Cooling algorithms~\cite{ber} are designed to find a solution for the
equations of motion associated to a local minimum of the energy
functional. It is relatively easy to write down the lattice equations of
motion. In particular it should be noted that the energy functional depends
linearly on the links. One finds
\be
\partial_{U_\mu(x)}\cE\propto U_\mu(x)\tilde U^\dagger_\mu(x)\!-\!\tilde U_\mu
(x)U^\dagger_\mu(x)=0,\quad\partial_{\rho(x)}\cE\propto\rho(x)\Bigl\{1\!-\!2
\kappa\!+\!2\lambda(\rho^2(x)\!-\!1)\Bigr\}\!-\!\tilde\rho(x)=0,
\label{eq:eqmo}
\ee
where
\be
\tilde{U}_\mu(x)=\half\kappa\rho(x)\rho(x\!+\!\hat\mu)+\tilde{U}_\mu(x;0),\quad
\tilde{U}_\mu(x;0)=\sum_{\nu\neq\mu}\left(\stapup+\stapdw\right),
\label{eq:Util}
\ee
\be
\tilde\rho(x)=\half\kappa\sum_\mu \rho(x\!+\!\hat\mu)\Tr(U_\mu(x))
+\rho(x\!-\!\hat\mu)\Tr(U_\mu(x\!-\!\hat\mu)).\label{eq:rtil}
\ee
The equations of motion for the links are solved by
\be
U_\mu(x)=\pm\tilde{U}_\mu(x)/\|\tilde{U}_\mu(x)\|.\label{eq:Ssol}
\ee
The positive sign is to be taken in order for the solution to have a smooth
continuum limit. The solution for the scalar field $\rho(x)$ is given by
the root of a cubic polynomial. If $1\!-\!2\kappa\!-\!2\lambda\!\geq\!0$,
equivalent to the condition $(aM_H)^2\!\leq\!12$, there is only one real root,
$\rho(x)=\rho_s(\tilde\rho(x))$, where
\be
\rho_s(\tilde\rho)\equiv\left[\frac{\tilde\rho}{4\lambda}\!+\!\sqrt{\left(
\frac{1\!-\!2\kappa\!-\!2\lambda}{6\lambda}\right)^3\!+\!\frac{\tilde\rho^2}{16
\lambda^2}}\
\right]^\third\!+\left[\frac{\tilde\rho}{4\lambda}\!-\!\sqrt{\left(
\frac{1\!-\!2\kappa\!-\!2\lambda}{6\lambda}\right)^3\!+\!\frac{\tilde\rho^2}{16
\lambda^2}}\ \right]^\third.\label{eq:rsol}
\ee
Cooling is performed by iterating these equations, i.e. replacing the link
and the scalar field by the right-hand side of these equations, sweeping
in a particular order through the lattice. With only nearest-neighbour
interactions, checkerboard-type updates are most efficient and allow for
vectorization of the algorithm. We use this cooling to first bring a random
configuration down to one that is smooth. But since the solutions we are
interested in have an unstable direction, we should switch to an algorithm
that does not make the solution decay along the unstable direction (to the
vacuum). This is achieved by taking the square of the equations of motion as
the minimizing functional~\cite{maw}, and devising an efficient algorithm for
minimization~\cite{us,lat94}. There are of course more sophisticated algorithms
to avoid decay along an unstable direction, but they tend to require
information on the Hessian of the energy functional, which is expensive
for large lattices.

\section{Saddle-point cooling}
We define $\hat S$ by summing the squares of the equations of motion,
$(\partial_{U_\mu(x)}\cE)^2$ and $(\partial_{\rho(x)}\cE)^2$,
\be
\hat S\hspace{-1mm}=\hspace{-1mm}\frac{1}{g^2a^3}\sum_{x,\mu}\Bigl\{
\Tr\!\left(\!\tilde{U}_\mu(x)\tilde{U}_\mu^{\dagger}(x)\!-\![U_\mu(x)
\tilde{U}_\mu^{\dagger}(x)]^2\!\right)\!+\!f\left(\rho(x)\left[1\!-\!2
\kappa\!+\!2\lambda(\rho^2(x)\!-\!1)\right]\!-\!\tilde\rho(x)\right)^2
\Bigr\},\label{eq:Sarj}
\ee
where $f$ is an arbitrary positive constant. One can show that in the
continuum limit
($D_\mu F_{\mu\nu}\equiv\partial_\mu F_{\mu\nu}\!+\![A_\mu,F_{\mu\nu}]$)
\be
\hat S=\frac{2}{g^2}\!\int d^3x|\Tr(D_\mu F_{\mu\nu}\!-\!\quart
A_\nu\phi^2)^2|\!+\!\frac{f\kappa}{2}\!\int d^3x \left[\partial_\mu^2\phi\!-\!
4\bar\lambda\phi^3\!+\!\half\Tr(A^2_\mu)\phi\!+\half\!M_H^2\phi\right]^2,
\ee
which has the dimension of $M_W^3$. Consequently, we will quote values of
$\hat S$ in units of $M_W^3/\alpha_W$. For $r$ finite, $\kappa$ has a non-zero
limit when $a\rightarrow 0$ (e.g. $\kappa(a=0,r=1)=0.1245$), we therefore took
$f=1$. Saddle-point cooling introduced in ref.~\cite{us} is designed to
minimize $\hat S$ down to its minimal value of zero. The value
of $\hat S$ is a direct measure for how close the cooled configuration is to
an exact lattice solution.

Finding an algorithm to minimize $\hat S$ is more complicated due to the
quadratic dependence on the link variables. It is not possible to analytically
find the minimum of $\hat S$ as a function of a single given link, keeping all
others (and $\rho(x)$) fixed. If $f=0$, where the scalar degree of freedom is
absent, the following algorithm~\cite{us,lat94} always lowers $\hat S$
\be
U^\prime_\mu(x)={M(U_\mu(x))-W_\mu(x)\over{\|M(U_\mu(x))-W_\mu(x)\|}}.
\label{eq:hSit}
\ee
We use the same algorithm here and add the prescription for updating the
scalar field. The definitions of $W_\mu(x)$ and $M(U_\mu(x))$ (specifying
the parts of $\hat S$ respectively linear and quadratic in $U_\mu(x)$)
will be split according to
\bea
M(U_\mu(x))&=&M^{(0)}(U_\mu(x))\!+\!M^{(1)}(U_\mu(x))\!+\!fM^{(2)}(U_\mu(x)),
\nonumber\\ W_\mu(x)&=&W^{(0)}_\mu(x)\!+\!W^{(1)}_\mu(x)\!+\!fW^{(2)}_\mu(x),
\eea
where the index 0 stands for the pure gauge part ($\kappa=0$, see
ref.~\cite{us}), the index 1 for the $\kappa$ dependent term arising through
the modified link equations of motion~\cite{lat94}, cmp.
eqs.~(\ref{eq:eqmo},\ref{eq:Util}),
and the index 2 stands for the part that comes from the scalar equations of
motion. Using the notation of $V_\mu^\alpha(x)$ for the $2(n\!-\!1)$ staples
in eq.~(\ref{eq:Util}), we have
\bea
M^{(0)}(U_\mu(x))+M^{(1)}(U_\mu(x))\hspace{-3mm}&\equiv&\hspace{-3mm}2\Tr\left(
U_\mu(x)\tilde{U}_\mu^\dagger(x)\right)\tilde{U}_\mu(x)+6\sum_\alpha\Tr\left(
U_\mu(x)V_\mu^\alpha(x)^\dagger\right)V_\mu^\alpha(x),\nonumber\\
M^{(2)}(U_\mu(x))\hspace{-3mm}&\equiv&\hspace{-3mm}-\frac{\kappa^2}{2}\left\{
\rho^2(x)+\rho^2(x\!+\!\hat\mu)\right\}\Tr\left(U_\mu(x)\right),
\eea
and
\bea
W^{(0)}_\mu(x)\hspace{-3mm}&=&\hspace{-3mm}2\sum_{\stackrel{a\neq-b}{a,b\neq
\pm\mu}}\wplaqone-\wplaqonem+2\sum_{\stackrel{a\neq-\mu}{b\neq\pm\mu,\pm a}}
\wplaqtwo-\wplaqtwom+\wplaqthree-\wplaqthreem\nonumber\\\nonumber\\
W^{(1)}_\mu(x)\hspace{-3mm}&=&\hspace{-3mm}\kappa\!\!\sum_{a\ne\pm\mu}\!\!\rho
(x\!+\!\hat\mu\!\!+\!\hat a)\Bigl\{\rho(x\!+\!\hat a)(\stapupap-\!\!\stapupaq\
)+\rho(x\!+\!\hat\mu)\stapupa\ (\linkap\!-\!\linkbp)\Bigr\}\!+\!\rho(x)\rho(
x\!+\!\hat a)\times\nonumber\\&&\hspace{8mm}(\linkb-\linka)\stapupa\ =\kappa
\!\!\sum_{a\ne\pm\mu}\!\!\rho(x\!+\!\hat\mu\!+\!\hat a)U_a(x)\Bigl\{\rho(x\!+\!
\hat a)(I\!\!-\!U^2_\mu(x\!+\!\hat a))U^\dagger_a(x\!+\!\hat\mu)+\nonumber
\\&&\hspace{5mm}\rho(x\!+\!\hat\mu)U_\mu(x\!+\!\hat a)(I\!\!-\!U^\dagger_a(
x\!+\!\hat\mu)^2)\Bigr\}\!+\!\rho(x)\rho(x\!+\!\hat a)(I\!\!-\!U_{a}^2
(x))U_{\mu}(x\!+\!\hat a)U_{a}^\dagger(x\!+\!\hat \mu)\nonumber\\\nonumber\\
W^{(2)}_\mu(x)\hspace{-3mm}&=&\hspace{-3mm}
\kappa\rho(x)\Bigl\{\frac{\kappa}{2}\sum_{a\neq-\mu}\rho(x\!+\!\hat\mu\!+
\!\hat a)\Tr(U_a(x\!+\!\hat\mu))\!-\!\rho(x\!+\!\hat\mu)(1\!-\!2\kappa\!+
\!2\lambda(\rho^2(x\!+\!\hat\mu)\!-\!1))\Bigr\}+\nonumber\\&&\hspace{5mm}
\kappa\rho(x+\hat\mu)\Bigl\{\frac{\kappa}{2}\sum_{a\neq\mu}\rho(x\!+\!\hat a)
\Tr(U_a(x))-\rho(x)(1\!-\!2\kappa\!+\!2\lambda(\rho^2(x)\!-\!1))\Bigr\},
\eea
with the unit vectors $\hat a,\hat b\in\{\pm\hat 1,\cdots,\pm\hat n\}$, and the
convention $U_{-a}(x)\equiv U_a^\dagger(x-\hat a)$. We only give the
explicit form for $W_\mu^{(1)}(x)$ and $W_\mu^{(2)}(x)$, referring for
$W^{(0)}_\mu(x)$ to eq.~(19) of ref.~\cite{us}. To implement this algorithm it
is useful to point out that $W^{(0)}_\mu(x)$ can be obtained by a sum over all
links in each staple of $\tilde U_\mu(x;0)$ (see eq.~(\ref{eq:Util})), with
each link $U_\ell$ replaced by the sum over $2U_\ell(U_P^\dagger-U_P)$, where
$U_P$ are plaquettes that end at this particular link, not overlapping with
the original staple. Likewise, $W^{(1)}_\mu(x)$ can be obtained as a sum over
all links in each staple of $\tilde U_\mu(x;0)$, with each link $U_\ell(y)$
replaced by $\kappa\rho(y)\rho(y+\hat\ell)(I-U^2_\ell(y))$. Alternatively,
one can describe $W_\mu^{(0)}(x)+W_\mu^{(1)}(x)$ by summing over all links in
each staple of $\tilde U_\mu(x;0)$, replacing each link $U_\ell(y)$ with
$2U_\ell(y)\{U^\dagger_\ell(y)\tilde U^\prime_\ell(y)-[U^\dagger_\ell(y)\tilde
U^\prime_\ell(y)]^\dagger\}$, where $\tilde U^\prime_\ell(y)$ is defined as
$\tilde U_\ell(y)$ in eq.~(\ref{eq:Util}), deleting in its sum over staples
the one staple that will have a link in common with the link $(x,x+\hat\mu)$.
For infinite Higgs self-coupling one
puts $\rho(x)\equiv1$, and $f=0$ to obtain the algorithm of ref.~\cite{lat94}.
This is consistent with the fact that $W_\mu^{(2)}(x)\!-\!M^{(2)}(U_\mu(x))
\propto\kappa (\rho(x)\partial_{\rho(x\!+\!\hat\mu)}\cE\!+\!\rho(x\!+\!\hat\mu)
\partial_{\rho(x)}\cE)$ vanishes when the scalar equations of motion are
enforced. Note that accidentally ref.~\cite{lat94} only listed the last of the
three terms in $W_\mu^{(1)}(x)$.

To verify the convergence of this part of the algorithm, we note that
$\hat S$ changes by the following exact amount~\cite{us}
\be
\delta\hat S(U_\mu(x))=-\frac{1}{2a^3g^2}\Tr\left(\delta U^\dagger_\mu(x)
\Bigl\{\|M(U_\mu(x))-W_\mu(x)\|\delta U_\mu(x)+M(\delta
U_\mu(x))\Bigr\}\right).
\ee
For finite values of $\lambda$, $M(U_\mu(x))$ is no longer positive.
Nevertheless, for $f=1$ and smooth configurations (near the continuum limit)
one easily sees that $M^{(2)}(U_\mu(x))$ scales to zero, and $\delta
\hat S\approx -112\|\delta U\|^2/(g^2a^3)$, see ref.~\cite{us} (below
eq.~(24)).

To complete the description of the algorithm for the general case, we have
to specify how to update the scalar field. We found that the ordinary cooling,
where we replace $\rho(x)$ by $\rho_s(\tilde\rho(x))$ (eq.~(\ref{eq:rsol}))
worked well. The apparent reason is that the unstable mode is dominated by
the gauge part of the energy functional. For large values of $M_H$ this is
no longer expected to be the case. We have also devised an updating of the
scalar field that is guaranteed to lower $\hat S$. Considering only the
part $\hat S_{\rho(x)}$ that depends on $\rho(x)$, we find up to irrelevant
constant factors,
\be
\hat S_{\rho(x)}=\left\{\rho(x)\left[1\!-\!2\kappa\!+\!2
\lambda(\rho^2(x)\!-\!1)\right]\!-\!\tilde\rho(x)\right\}^2+B(x)\rho(x)
+C(x)\rho^2(x),\label{eq:Srho}
\ee
where
\bea
B(x)&=&\kappa\sum_a\Bigl\{f^{-1}\!\rho(x\!+\!\hat a)\Tr[(I\!-\!U^2_a(x))
\tilde U_a^\dagger(x;0)]\!-\!\Bigl[\rho(x\!+\!\hat
a)(1\!-\!2\kappa\!+\!2\lambda
(\rho^2(x\!+\!\hat a)\!-\!1))\nonumber\\&&\hspace{4cm}-\frac{\kappa}{2}
\sum_{b\neq-a}\rho(x\!+\!\hat a\!+\!\hat b)\Tr(U_b(x\!+\!\hat a))\Bigr]
\Tr(U_a(x))\Bigr\}
\eea
and
\be
C(x)=\frac{\kappa^2}{4}\sum_a\Bigr\{f^{-1}\!\rho^2(x\!+\!\hat a)\Tr(I\!\!-\!
U_a^2(x))+\Tr^2(U_a(x))\Bigr\}.
\ee
As before we take $\hat a,\hat b\in\{\pm\hat 1,\cdots,\pm\hat n\}$ and use the
convention that $U_{-a}(x)\equiv U_a^\dagger(x\!-\!\hat a)$
and $\tilde U_{-a}(x;0)\equiv\tilde U_a^\dagger(x\!-\!\hat a;0)$.
Note that $\hat S$ is a sixth order polynomial in $\rho(x)$. We will show
that under very mild conditions $\hat S$ is a convex function of $\rho(x)$.
This greatly simplifies the problem of minimizing $\hat S$ with respect to
$\rho(x)$, using ordinary Newton-Raphson. Provided $(aM_H)^2\!\leq\!12$,
the second derivative of $\hat S$ with respect to $\rho(x)$ has a unique
minimum at $\rho_m(x)\equiv\sqrt{\frac{2}{5}}\rho_s(\sqrt{\frac{5}{32}}
\tilde\rho(x))$, with $\rho_s$ defined as in eq.~(\ref{eq:rsol}). At this
minimum
\be
\left\{\partial^2_{\rho(x)}\!\hat S\right\}_{\rm
min}\!=\frac{16(1\!-\!2\kappa\!
-\!2\lambda)^3\!-\!27\lambda\tilde\rho^2(x)+3\lambda\left[8(1\!-\!2\kappa\!-\!
2\lambda)\rho_m(x)\!-\!3\tilde\rho(x)\right]^2}{8(1\!-\!2\kappa\!-\!2\lambda)}
+\!2C(x).
\ee
As $C(x)\!\geq\!0$, this is always positive provided
$27\lambda\tilde\rho^2(x)\!<\!16(1\!-\!2\kappa\!-\!2\lambda)^3$, or
\be
\left(\frac{\tilde\rho(x)}{6\kappa v}\right)^2<\frac{128(1-(aM_H)^2/12)^3
}{9(aM_H)^2}.\label{eq:convex}
\ee
Since $\tilde\rho(x)/(6\kappa v)\!\leq\!\bar\rho(x)/v$, where $\bar\rho(x)$ is
the average over the nearest neighbours, we conclude that in all practical
cases
$\hat S$ is indeed a convex function of $\rho(x)$. The unique minimum of
eq.~(\ref{eq:Srho}) is rapidly found by the iteration
\be
\rho^\prime(x)=\rho(x)-s \partial_{\rho(x)}\!\,\hat S/\partial^2_{\rho(x)}
\!\,\hat S,
\ee
where $s$ is a free parameter used to speed up the algorithm (the standard
value being $s=1$). The convexity guarantees that $\hat S$ is always
strictly lowered, unless $\rho(x)$ is already at its minimum, like for
eq.~(\ref{eq:hSit}). For each sweep one performs both iterations only
once for each site (one does not gain speed by multiple iterations per
site, as the convergence of the algorithm is determined by the lowest
eigenvalue of the square of the Hessian of the energy functional~\cite{us}).

Although the algorithm may seem difficult to implement, its main
advantage is that it is deterministic, with a good understanding of its
convergence~\cite{us}. Most importantly, the stringent tests that $\hat S$
must always decrease under saddle-point cooling, and the condition that for
a solution $\hat S$ must vanish to a high degree of accuracy, are guarantees
that the algorithm was programmed correctly. Also the test for convexity of
$\hat S$ was never seen to be violated after initial ordinary cooling.
Testing the algorithm without this initial cooling is, even in the absence
of the scalar field, not very useful as it tends to get trapped in dislocations
when starting from a random configuration. This is avoided by ordinary cooling
due to the choice of positive sign in eq.~(\ref{eq:Ssol}).

\section{Finite size scaling}

To obtain infinite volume results in the continuum one needs to first
extrapolate at a fixed volume $LM_W=N\sqrt{\bar\kappa/2}$ to the continuum
by taking the limit $\bar\kappa=2(aM_W)^2\rightarrow0$, which is achieved by
fitting to
\be
\cE(M_WL,\bar\kappa)=\cE_{\rm sph}(M_WL)+\cE_1(M_WL)\bar\kappa/2+\cE_2(M_WL)
\bar\kappa^2/4+\cdots.\label{eq:afit}
\ee
For small enough lattice spacings this extrapolation can be done accurately.
Subsequently one extrapolates these continuum results to an infinite volume.
The more information one has available on the asymptotic behaviour of $\cE(L)$
the more accurate one can extract $\cE^\infty_{\rm
sph}\equiv\cE_{sph}(\infty)$.
Introducing the shifted field $\fie=\phi-(8\bar\lambda)^{-\half}M_H$,
we denote by $(\bar A,\bar\fie)$ the infinite volume solution~\cite{yaf} and
by $(\delta_L A,\delta_L\fie)$ the correction due to the periodic boundary
conditions. The linearized equations of motion are those of non-interacting
massive vector and scalar fields. For the vector field the linearized equations
of motion impose $\partial_i A_i^a(x)=0$ and the most general rotationally
covariant solutions are given by
\bea
\cA_i^a(x)&\equiv&C_W\Bigl\{\cos(\delta)\veps_{iaj}
\partial_j K(rM_W)+\sin(\delta)M_W^{-1}\veps_{ibj}\veps_{abk}\partial_j
\partial_k K(rM_W)\Bigr\}/\sqrt{\alpha_W},\nonumber\\ \Phi(x)&\equiv&C_H
M_H K(rM_H)/\sqrt{\alpha_W},\quad K(r)\equiv\frac{\exp(-r)}{r},\label{eq:free}
\eea
where $\delta$ is non-zero for the deformed sphalerons~\cite{yaf}
($M_H>12M_W$) and zero for the ordinary sphalerons ($M_H<12M_W$).
These functions describe the solution $(\bar A,\bar\fie)$ at large distances
$r\equiv\|x\|\rightarrow\infty$.
At distances $\half L\geq R\gg M^{-1}$ from the center of the solution, the
fields satisfy the linearized equations of motion up to {\em relative} errors
of the order of $e^{-MR}$, where $M$ is the smallest of the
two masses in the problem. In this region the solution can be described
by periodic copies
\be
(A(x),\fie(x))=\sum_{\vec n\in\zahlen^3}(\cA(x\!+\!\vec n L)
,\Phi(x\!+\!\vec n L)).
\label{eq:copies}
\ee
We will now split the energy density $V(\bar A+\delta_L A,\bar\fie
+\delta_L\fie)$ into $V(\bar A,\bar\fie)$ and terms linear and quadratic
in the shifted fields. Higher order terms are suppressed to $\cO(e^{-3ML/2})$.
To this order the term quadratic in the shifted fields, sums with the
zeroth order term to the energy of the sphaleron in an infinite volume,
after integration over the periodic box. This is because the dominating
contribution for the quadratic term comes from the region near the boundary
of the periodic box where one can neglect the interactions between the copies.
To $\cO(e^{-3ML/2})$ all volume dependence is therefore determined by the
term linear in the shift of the fields, for which we can use the
$(\bar A,\bar \fie)$ equations of motion, leaving only a boundary term
\be
\cE_{\rm sph}(L)=\cE^\infty_{\rm sph}+\int_{-L/2}^{L/2}d^3x~\partial_j(\delta_L
A_i^a(x)\bar F_{ji}^a(x)+\delta_L\fie\partial_j\bar\fie)+\cO(e^{-3ML/2}).
\ee
The surface integral is evaluated using eq.~(\ref{eq:copies}), together
with the explicit expressions of eq.~(\ref{eq:free}). Each of the six
faces of the cube gives the same contribution to the surface integral.
We extend the integral over one face to the whole plane, at the expense of
an error of $\cO(e^{-\sqrt{2}ML})$. To this order only the nearest copy
will contribute and we can ignore the non-linear term in the expression
for the field strength. With $y=x\!-\!\hat 1\!L$, one has
\bea
\cE_{\rm sph}(L)&=&\cE^\infty_{\rm sph}+\delta\cE(L)+\cO(e^{-\sqrt{2}ML}),
\\ \delta\cE(2x_1)&\equiv&\frac{6}{\alpha_W}\int dx_2dx_3 ~\Bigl\{
\cA_i^a(y)(\partial_1\cA^a_i(x)\!-\!\partial_i\cA_1^a(x))\!+
\!\Phi(y)\partial_1\Phi(x)\Bigr\}.\nonumber
\eea
Using $\partial_i^2K(rM)=M^2K(rM)$, the integrand between curly brackets can
be simplified to
\bea
&&C_W^2\Bigl\{\cos^2(\delta)\partial_iK(y)[\delta_{i1}\partial_k^2+
\partial_i\partial_1]K(x)+\sin^2(\delta)\partial_iK(x)[\delta_{i1}\partial_k^2
+\partial_i\partial_1]K(y)+\\&&\hspace{2mm}\half\veps_{i1k}\sin(2\delta)[
M_W\partial_iK(y)\partial_kK(x)-\partial_i\partial_aK(y)\partial_k
\partial_aK(x)/M_W]\Bigr\}+C_H^2M_H^2K(y)\partial_1K(x).\nonumber
\eea
Performing the surface integral one easily sees that the term proportional to
$\sin(2\delta)$ is a total derivative with respect to $x_2$ and
$x_3$, whereas at $x_1=\half L$ the other terms reduce after some partial
integrations to ($K^\prime(r)\equiv dK(r)/dr$)
\be
\delta\cE(L)\!=\!\frac{3L}{\alpha_W}\int\!\frac{dx_2dx_3}{r}\Bigl\{C_H^2
M_H^3K(rM_H)K^\prime(rM_H)-2\cos(2\delta)C_W^2M_W^3K^\prime(rM_W)K(rM_W)\Bigr\}.
\ee
With $r^2=\quart L^2+x_2^2+x_3^2\equiv\quart L^2+s^2$ and $sds=rdr$, and the
fact that the integrand is a total derivative in $r$, we get
the following exact result
\be
\delta\cE(L)=24\pi\cos(2\delta)C_W^2\frac{M_W}{\alpha_W}\frac{\exp(-M_WL)}{
M_WL}-12\pi C_H^2\frac{M_H}{\alpha_W}\frac{\exp(-M_HL)}{M_HL}.\label{eq:Lfit}
\ee
The dimensionless constants $\delta$, $C_W$ and $C_H$ are expected to depend on
the ratio $M_H/M_W$. We have thus found the remarkable result that subleading
corrections are not powerlike (as was assumed in ref.~\cite{lat94}), but
exponential. With the help of these asymptotic expansions we will be able
to extract $\cE^\infty_{\rm sph}$ to rather high accuracy from our data.

\section{Results}

As is usual in lattice gauge theories, or for that matter any discretization
technique, there are two conflicting sources of numerical errors. On the one
hand the correlation length ($1/M$) should be much larger than the lattice
spacing to minimize lattice artefacts, on the other hand it should be much
smaller than $L=aN$ to minimize finite size errors.

For small values of $aM_W$ the electroweak sphaleron tends to develop
additional unstable modes. There are two reasons due to finite volume
effects. The first reason is that the rotational invariance will only be
approximate such that the energy functional will no longer be flat as a
function of the rotational moduli. As saddle-point cooling works irrespective
of the number of unstable modes, the solution might be attracted to a saddle
point with additional (usually small) negative eigenvalues of the Hessian.
Secondly, the saddle point associated to the pure gauge finite volume
sphaleron~\cite{us}, obtained by putting $\kappa=0$, will be lighter
than the electroweak sphaleron for small volumes. At finite values of
the Higgs self-coupling the pure gauge finite volume sphaleron remains an
exact solution by putting $\rho(x)=0$. It has an energy $72.605/(g^2L)+
\half M_W^2M_H^2L^3/g^2$. At infinite Higgs self-coupling the solution will
be deformed (we have no freedom to choose $\rho(x)=0$ to make the gauge field
massless). In this case the crossing occurs at $M_WL\approx 2.5$.
We observed below the crossing of these {\em distinct} solutions
that the electroweak sphaleron acquires additional unstable modes. (The other
saddle point acquires extra unstable modes for larger volumes. Close inspection
reveals that the changes do {\em not} occur exactly at the crossing.)

For large values of $aM_W$ both translational and rotational invariance will
be broken by the coarseness of the lattice. This will cause the energy
functional to develop spurious saddle points and one might get trapped
in one with additional negative modes, as for the breakdown of rotational
invariance due to a finite volume. We typically will choose $aM_W$ such
that the eigenvalues of the Hessian associated to the approximate zero
modes are not too big. For finite values of the Higgs
self-coupling another feature will cause problems at large values of
$aM_W$, associated to an enhanced gauge symmetry of the solution.
In the unitary gauge the energy functional is generally only invariant
under global gauge rotations. However, suppose that the exact lattice solution
will have $\rho(0)=0$, as is true in the continuum. It is then easily seen
that the hopping term of the energy functional is insensitive to all links
connected to the origin. The energy functional is therefore invariant under a
gauge transformation that is non-trivial at $x=0$ only, as this does not affect
the plaquette contribution to the energy. In particular at no expense in
energy one can flip the sign of the trace of the links connected to the origin.
In the way we prepared the configurations this will not occur if all
links are close to the identity. But at moderately large lattice spacing
or small volumes $\rho(0)$ is no longer exactly zero. The hopping term now
depends weakly on the gauge transformation
at the origin. This tends to favour a negative value of the trace for only
one of the links connected to the origin. (From this we also found solutions
with the trace of all links positive, with almost identical energies.)
Initially, a negative value for the trace of one of the links mislead us to
believe that we were dealing with dislocations.

Putting all constraints in we found for $M_H=\infty$ the
window of allowed values to be $M_WL\geq2.5,\ aM_W\leq0.40$, for $M_H=M_W$
the window is $M_WL\geq3.8,\ aM_W\leq0.60$ and for $M_H=\tfour M_W$ it is
$M_WL\geq4.0,\ aM_W\leq0.65$.

Figure 1 gives the energy density profiles of the electroweak sphaleron at
each of the three Higgs masses. We should not directly use eq.~(\ref{eq:Swil}),
but first average over all directions of the links connected to a point $x$
(without affecting the total energy), in order to compute the energy density at
this point. Note that for $M_H=\infty$ the solution is very much more peaked
in the core region and will have larger lattice artefacts. The behaviour in
the tail region is similar to the case where $M_W=M_H$. For $M_H=\tfour M_W$
this tail region is dominated by the decay of the scalar field. Also plotted
in figure 1 is the behaviour of $\rho(x)/v$ for $M_H=M_W$ and $M_H=\tfour M_W$
at $M_WL=4$. Because of finite volume effects the scalar field does not exactly
equal its expectation value at the boundary.
Likewise it does not quite go to zero at the center, which is also due to
finite lattice spacing errors.
\begin{figure}[htb]
\vspace{2.4cm}
\includegraphics{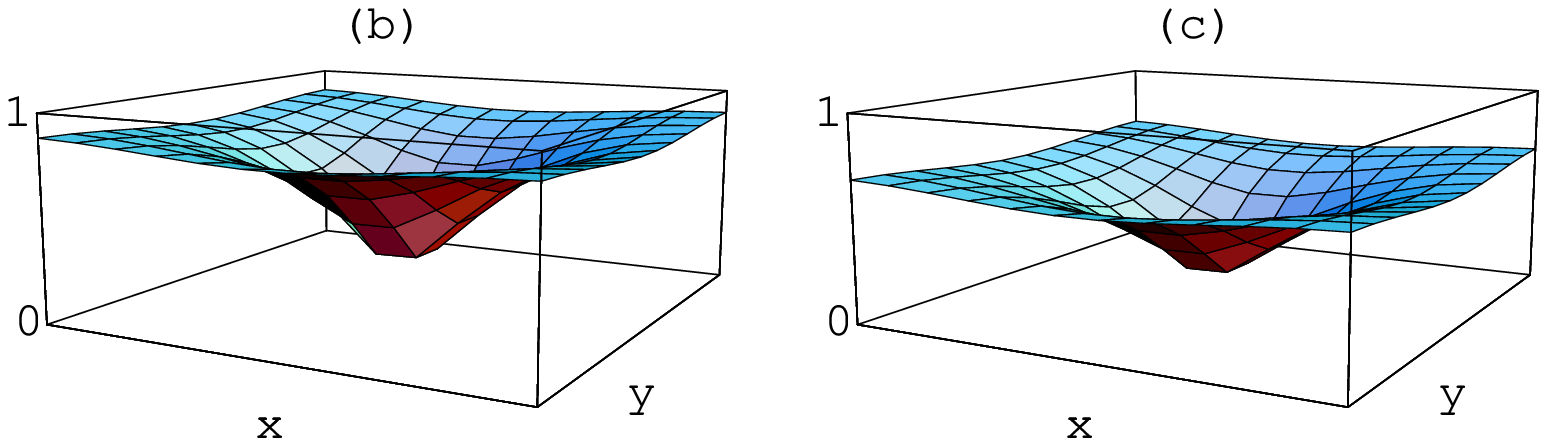}
\end{figure}
\begin{figure}[htb]
\vspace{4.5cm}
\includegraphics{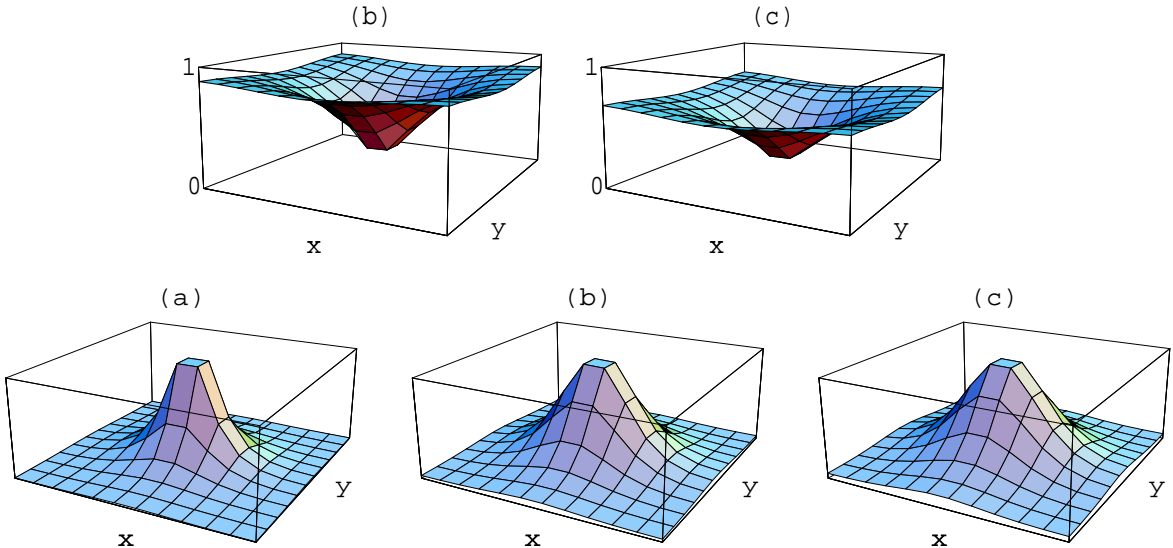}
\caption{
The scalar field (top) and the energy density (bottom) in a plane
through the center of the electroweak sphalerons for $(a):M_H=\infty$ at
$M_WL=2.53$, $(b):M_H=M_W$ and $(c): M_H=\tfour M_W$, both at $M_WL=4.0$.
The energy density is normalized to its peak value (respectively $0.093$,
$0.025$ and $0.016 M^4_W/\alpha_W$) and the scalar field $\rho$ to its
expectation value $v$.}
\label{fig:profile}
\end{figure}

The way we obtained the required
configurations was by first constructing a sphaleron for the frozen-length
Higgs model, starting at $N=8$. All links at the boundary were first put to the
identity, which serves the purpose of positioning the solution in the center
of the lattice and of lifting the energy of the finite volume sphaleron by a
considerable amount. The latter helps avoid getting trapped in that solution.
Centering the energy profile will reduce the probability of getting stuck in a
saddle point with spurious unstable modes due to the breakdown of translational
and rotational invariance. We then release the frozen boundary condition and
compute the Hessian after cooling to verify that we have one unstable mode
only. This way the maximal energy density occurs at the center of a
plaquette, see fig.~1. The solutions where the maximum occurs at a lattice
point are higher in energy.  One can now change the lattice spacing in small
steps to scan the desired range of parameters. For $N=12$ and 16 the initial
configurations were generated from the one at $N=8$, by embedding it in the
large lattice (links parallel to the boundary remain constant and those
perpendicular to the boundary are put to unity) and adjusting the lattice
spacing. For $M_W=M_H$ we generated the sphalerons for $N=8$ from the
frozen-length sphaleron (in not too small a volume) by adding the scalar
field, set to its expectation value $v$. Varying the lattice spacing in small
steps allows one again to scan the desired range of parameters. Finally, the
sphalerons with $M_H=0.75M_W$ were generated from the ones with $M_W=M_H$
by simply adjusting the parameters.

\begin{table}[h]
\setlength{\tabcolsep}{.44pc}
\newlength{\digitwidth} \settowidth{\digitwidth}{\rm 0}
\catcode`?=\active \def?{\kern\digitwidth}
\label{tab:sval1}
\begin{tabular}{@{}c@{\extracolsep{\fill}}cccccccr}
\hline
           \multicolumn{8}{l}{\vspace{-4.5mm}}\\
           \multicolumn{1}{c}{$M_H=\infty$}
           &\multicolumn{3}{c} {$\frac{{\cal E}}{M_W/\alpha_W}$}
           &\multicolumn{1}{c}{$-\frac{\omega^2}{M_W^2}$}\\
\cline{2-4}\cline{5-5}
           \multicolumn{1}{c}{$LM_W$}
           & \multicolumn{1}{c} {$N=8$}
           & \multicolumn{1}{c} {$N=12$}
           & \multicolumn{1}{c} {$N=16$}
           & \multicolumn{1}{c} {$N=8$}
           &\multicolumn{1}{c}{$\frac{{\cal E}_{\rm sph}}{M_W/\alpha_W}$}
           &\multicolumn{1}{c}{$\ \frac{{\cal E}_{1}}{M_W/\alpha_W}$}
           &\multicolumn{1}{c}{$\ \frac{{\cal E}_{2}}{M_W/\alpha_W}$}
           \\
\hline
$ 2.5298 $&$  5.2041(2) $&$   5.4153(3) $&$ 5.4699(4) $&$ 5.846 $&
$ 5.525 $&$ -1.89 $&$ -13.2 $\\
$ 2.7713 $&$  5.0117(1) $&$   5.2598(2) $&$ 5.3258(4) $&$ 5.442 $&
$ 5.395 $&$ -2.00 $&$  -9.9 $\\
$ 2.8823 $&$  4.9352(1) $&$   5.2001(3) $&$ 5.2728(4) $&$ 5.325 $&
$ 5.351 $&$ -2.15 $&$  -8.1 $\\
$ 2.9933 $&$  4.8645(1) $&$   5.1459(3) $&$ 5.2263(5) $&$ 5.250 $&
$ 5.316 $&$ -2.34 $&$  -6.3 $\\
$ 3.2000 $&$  4.7446(1) $&$   5.0549(4) $&$ 5.1535(5) $&$ 5.231 $&
$ 5.273 $&$ -2.87 $&$  -2.7 $\\
\hline
\multicolumn{8}{l}{\vspace{-4mm}}\\
\multicolumn{8}{l}{$
\frac{\cE_{\rm sph}(L)}{M_W/\alpha_W}=5.09(1)+13.6(5)
\frac{e^{-M_WL}}{M_WL},\quad \frac{\cE_v}{M_W/\alpha_W}=5.0707
$}\\
\multicolumn{8}{l}{}\\
\hline
           \multicolumn{8}{l}{\vspace{-4.5mm}}\\
           \multicolumn{1}{c}{$M_H=M_W$}
           &\multicolumn{3}{c} {$\frac{\cE}{M_W/\alpha_W}$}
           &\multicolumn{1}{c}{$-\frac{\omega^2}{M_W^2}$} \\
\cline{2-4}\cline{5-5}
           \multicolumn{1}{c}{$LM_W$}
           & \multicolumn{1}{c} {$N=8$}
           & \multicolumn{1}{c} {$N=12$}
           & \multicolumn{1}{c} {$N=16$}
           & \multicolumn{1}{c} {$N=8$}
           &\multicolumn{1}{c}{$\frac{{\cal E}_{\rm sph}}{M_W/\alpha_W}$}
           &\multicolumn{1}{c}{$\ \frac{{\cal E}_{1}}{M_W/\alpha_W}$}
           &\multicolumn{1}{c}{$\ \frac{{\cal E}_{2}}{M_W/\alpha_W}$}
           \\
\hline
$ 3.8000  $&$  3.6564(1) $&$  3.7090(1) $&$ 3.7261(8) $&$ 2.371$&
$ 3.747 $&$ -0.36 $&$ -0.18 $\\
$ 4.0000  $&$  3.6249(1) $&$  3.6830(1) $&$ 3.7013(5) $&$ 2.313$&
$ 3.723 $&$ -0.34 $&$ -0.22 $\\
$ 4.1600  $&$  3.6026(1) $&$  3.6657(1) $&$ 3.6852(4) $&$ 2.287$&
$ 3.708 $&$ -0.33 $&$ -0.24 $\\
$ 4.2208  $&$  3.5946(1) $&$  3.6597(2) $&$ 3.6798(5) $&$ 2.281$&
$ 3.704 $&$ -0.33 $&$ -0.24 $\\
$ 4.4000  $&$  3.5724(1) $&$  3.6440(1) $&$ 3.6660(5) $&$ 2.280$&
$ 3.692 $&$ -0.33 $&$ -0.23 $\\
$ 4.6000  $&$  3.5490(1) $&$  3.6290(1) $&$ 3.6527(4) $&$ 2.308$&
$ 3.680 $&$ -0.31 $&$ -0.27 $\\
$ 4.8000  $&$  3.5258(1) $&$  3.6160(1) $&$ 3.6418(4) $&$ 2.380$&
$ 3.671 $&$ -0.29 $&$ -0.30 $\\
\hline
\multicolumn{8}{l}{\vspace{-4mm}}\\
\multicolumn{8}{l}{$
\frac{\cE_{\rm sph}(L)}{M_W/\alpha_W}=3.6406(6)+18.1(2)
\frac{e^{-M_WL}}{M_WL},\quad \frac{\cE_v}{M_W/\alpha_W}=3.6417
$}\\
\multicolumn{8}{l}{}\\
\hline
           \multicolumn{8}{l}{\vspace{-4.5mm}}\\
           \multicolumn{1}{c}{$M_H=\tfour M_W$}
           &\multicolumn{3}{c} {$\frac{\cE}{M_W/\alpha_W}$}
           &\multicolumn{1}{c}{$-\frac{\omega^2}{M_W^2}$} \\
\cline{2-4}\cline{5-5}
           \multicolumn{1}{c}{$LM_W$}
           & \multicolumn{1}{c} {$N=8$}
           & \multicolumn{1}{c} {$N=12$}
           & \multicolumn{1}{c} {$N=16$}
           & \multicolumn{1}{c} {$N=8$}
           &\multicolumn{1}{c}{$\frac{{\cal E}_{\rm sph}}{M_W/\alpha_W}$}
           &\multicolumn{1}{c}{$\ \frac{{\cal E}_{1}}{M_W/\alpha_W}$}
           &\multicolumn{1}{c}{$\ \frac{{\cal E}_{2}}{M_W/\alpha_W}$}
           \\
\hline
$ 4.0000 $&$  3.4193(2) $&$  3.4578(2) $&$ 3.4703(4) $&$ 1.916$&
$ 3.486 $&$ -0.24 $&$ -0.11 $\\
$ 4.4000 $&$  3.4078(2) $&$  3.4585(2) $&$ 3.4743(3) $&$ 1.886$&
$ 3.493 $&$ -0.24 $&$ -0.15 $\\
$ 4.8000 $&$  3.3925(1) $&$  3.4584(1) $&$ 3.4782(3) $&$ 1.934$&
$ 3.501 $&$ -0.24 $&$ -0.17 $\\
$ 5.2000 $&$  3.3699(2) $&$  3.4565(2) $&$ 3.4807(3) $&$ 2.100$&
$ 3.507 $&$ -0.23 $&$ -0.23 $\\
\hline
\multicolumn{8}{l}{\vspace{-4mm}}\\
\multicolumn{8}{l}{$
\frac{\cE_{\rm sph}(L)}{M_W/\alpha_W}=3.530(3)+24(4)\frac{e^{-M_WL}}{M_WL}-
12(2)\frac{e^{-M_HL}}{M_WL},\quad\frac{\cE_v}{M_W/\alpha_W}=3.5355
$}\\
\multicolumn{8}{l}{\vspace{-4mm}}\\
\hline
\end{tabular}
\caption{Lattice data for the sphaleron energies $\cE$, the negative
eigenvalue of the Hessian on a $8^3$ lattice, the fit to the lattice
spacing dependence, and volume dependence. We give as many digits
as we believe to be significant. The variational result is
denoted by $\cE_v$.}
\end{table}

In table 1 we present the results for the sphaleron energies, the negative
eigenvalue for the $N=8$ Hessian, the fit to the lattice spacing dependence
(eq.~(\ref{eq:afit})) and to the volume dependence (eq.~(\ref{eq:Lfit})).
We list the variational results~\cite{yaf} for $\cE_{\rm sph}^\infty$ as
$\cE_v$. For the frozen-length Higgs model~\cite{lat94} we have here performed
some further cooling down to $\hat S<10^{-5}M^3_W/\alpha_W$ for $N=16$ (and
one or two orders of magnitude smaller for $N=8$ and $12$), to justify the
five digit accuracy (estimated errors in the last digit given between
brackets). As was to be expected, one finds appreciable lattice artefacts
for the case $M_H=\infty$. On the other hand they are comfortably small for
$M_H\approx M_W$.
To demonstrate this further we also computed at $M_W=M_H$ and $N=8$ the
energies for $aM_W=0.644,\,0.663,\,0.788$ and $0.825$ giving respectively
$\cE=3.493,\,3.476,\,3.355$ and $3.305M_W/\alpha_W$. These are solutions with
a negative trace for one of the links, as described above, which is why we
did not use these values for the finite size scaling. Nevertheless, it shows
that even for these rather coarse lattices, the error in the sphaleron energy
is only 10\%, which was somewhat surprising. For these solutions the lattice
artefacts are described well by the fit to the lattice spacing dependence
given in the table (at $M_WL=4.8$ for $M_W=M_H$).

\begin{figure}[htb]
\vspace{5.2cm}
\includegraphics{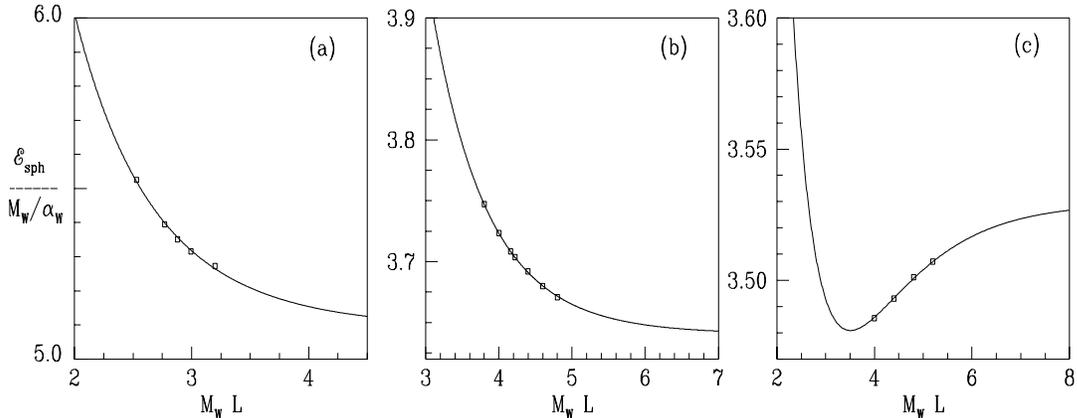}
\caption{Continuum extrapolated values for ${\cal E}_{\rm sph}$
as a function of the physical volume $M_W L$, combined with fits to the finite
volume behaviour for $(a):M_H=\infty$, $(b):M_H=M_W$ and $(c):M_H=\tfour M_W$.}
\label{fig:vol}
\end{figure}

Figure 2 compares the
fit to eq.~(\ref{eq:Lfit}) with the continuum extrapolated lattice data.
Our results for $\cE_{\rm sph}^\infty$ are in very good agreement with the
variational analysis~\cite{yaf}, particularly for $M_W=M_H$, where we achieve
an accuracy of better than .05\%. For $M_H=\infty$, a much better fit with
$\cE_{\rm sph}^\infty=5.075(5)M_W/\alpha_W$ is obtained when dropping the
largest-volume data point. This was the only case where the energy is lowered
significantly as compared to ref.~\cite{lat94}, seemingly because we are
unable to avoid being trapped in saddle points with additional unstable modes
at $N\geq12$. Note that at some point subleading exponential corrections will
start to become relevant too. For $M_H=M_W$, dropping the last point gives
$\cE_{\rm sph}^\infty=3.6412(8)M_W/\alpha_W$, whereas for $M_H=\tfour M_W$
we find $3.535M_W/\alpha_W$. The values of $C_H^2$ and $\cos(2\delta)C_W^2$
obtained from these fits (cmp. table 1) agree with what one can roughly extract
from the figures of ref.~\cite{yaf}.

An alternative method for studying the electroweak sphaleron on the lattice
is being considered by Ambj\o rn and Krasnitz, using the Chern-Simons
functional to constrain the cooling~\cite{kra}. It has the advantage of
allowing ordinary cooling rather than saddle-point cooling and might also
be used for computing the energy along the tunnelling path. In the continuum
the sphaleron has a Chern-Simons number of a half (compared to the trivial
vacuum). Its disadvantage is that implementing the Chern-Simons functional on
a lattice, the electroweak sphaleron will only approximately be characterized
by a Chern-Simons number of exactly one half. As the tunnelling rate depends
exponentially on the sphaleron energy, our results might be particularly
useful in numerical checks of the semiclassical determination~\cite{mcl} of
the tunnelling rate at small temperatures~\cite{amb,smi} (for the Abelian Higgs
model in 1+1 dimensions see ref.~\cite{boc}), as our method gives the exact
saddle-point solution for the sphaleron on a lattice.

\section*{Acknowledgements}

This work was supported in part by FOM and by a grant from NCF for use
of the Cray C98. M.G.P. was also supported by  MEC. We are grateful to
Jan Smit and William Tang for useful discussions and to Alexander Krasnitz
for correspondence, explaining his results.

\end{document}